\documentclass[figures]{epl}
\usepackage{graphicx}
\usepackage{rcs}

\title{Delocalization and Heisenberg's uncertainty relation}
\shorttitle{Delocalization and Heisenberg's uncertainty relation}
\shortauthor{G.-L. Ingold \etal}
\author{%
Gert-Ludwig Ingold\thanks{E-mail: \email{Gert.Ingold@physik.uni-augsburg.de}}
\and
Andr{\'e} Wobst
\and
Christian Aulbach
\and
Peter H{\"a}nggi
}
\institute{Institut f{\"u}r Physik, Universit{\"a}t Augsburg, 
Universit{\"a}tsstra{\ss}e 1, D-86135 Augsburg,\\ Germany}

\begin{document}
\maketitle

\begin{abstract}
In the one-dimensional Anderson model the eigenstates are localized for
arbitrarily small amounts of disorder. In contrast, the Harper model
with its quasiperiodic potential shows a transition from extended to 
localized states. The difference between the two models becomes particularly 
apparent in phase space where Heisenberg's uncertainty relation imposes a 
finite resolution. Our analysis points to the relevance of the coupling 
between momentum eigenstates at weak potential strength for the delocalization 
of a quantum particle.
\end{abstract}

\section{Introduction}
The delocalization of two interacting quantum particles in a disordered
potential has been the subject of intensive research since it was first 
addressed \cite{shepe94,imry95,weinm95}. More recent work has generalized
the Harper or Aubry-Andr{\'e} model \cite{harpe55,aubry80} to study the
behaviour of two interacting particles in a quasiperiodic potential
\cite{barel96,shepe96,eilme01}. Here, the interaction rather leads
to a tendency towards localization. For finite densities, no clear indication
of an interaction dependence of the phase transition in the Harper model has
been found \cite{schus02}.

Already in the absence of interaction, a quantum particle in one dimension 
exhibits a very different behaviour depending on the potential in which it 
is moving. For a periodic potential, the eigenstates are extended Bloch 
waves characterized by a quasimomentum. On the other hand, already a small 
amount of disorder suffices to localize the particle \cite{ander58}. A 
different situation can arise when motion on a lattice is considered. Then, 
for a periodic potential incommensurate with the underlying lattice, a 
transition from delocalized to localized states occurs as the potential 
strength reaches a critical value. 

In order to gain more insight into the localization properties of a quantum 
particle in one dimension, we compare two lattice models, the Anderson model 
\cite{ander58} and the Harper or Aubry-Andr\'e model \cite{harpe55,aubry80}. 
An analysis of the phase space properties of the energy eigenstates will reveal 
the dependence on the nature of the coupling between momentum eigenstates due 
to the potential in which the particle is moving.

\section{Random and quasiperiodic potential}
The Hamiltonian of the Anderson model is given by \cite{ander58}
\begin{equation}
H=-\sum_n\left(\vert n\rangle\langle n+1\vert + \vert n+1\rangle\langle n\vert
\right) + W\sum_n v_n\vert n\rangle\langle n\vert
\label{eq:ha}
\end{equation}
where the Wannier states $\vert n\rangle$ are localized at the sites 
$n=1,\dots,L$ of a ring with periodic boundary conditions. The first term on
the right-hand side describes the kinetic energy which defines the energy
scale. The random potential of strength $W$ is expressed by the second term. 
The distribution of on-site energies is determined by the coefficients $v_n$ 
distributed uniformly on the interval $[-1/2;1/2]$. In the limit $L\to\infty$ 
the eigenstates of the Anderson model are known to localize for any 
nonvanishing potential strength $W$ \cite{ander58}.

The second model of interest, the Harper model, is defined by the Hamiltonian
\cite{harpe55,aubry80}
\begin{equation}
H=\sum_n\left(\vert n\rangle\langle n+1\vert + \vert n+1\rangle\langle n\vert
\right) + \lambda\sum_n\cos\left(2\pi\beta n\right)
\vert n\rangle\langle n\vert
\label{eq:hhx}
\end{equation}
where the random potential in (\ref{eq:ha}) has been replaced by a 
quasiperiodic potential if the parameter $\beta$ assumes an irrational number 
in the limit $L\to\infty$. Then, the nature of the eigenstates 
depends on the value of the parameter $\lambda$. For $\lambda<2$, all states 
are extended while they are localized for $\lambda>2$ \cite{aubry80}.

For finite size systems, it is convenient to choose $\beta=F_{i-1}/F_i$ where 
$F_{i-1}$ and $F_i$ are two successive Fibonacci numbers \cite{kohmo83}. In the 
limit of large systems $\beta$ approaches the inverse of the golden 
mean, $(\sqrt{5}-1)/2$. With this choice of $\beta$ the system contains $L=F_i$ 
lattice sites with $F_{i-1}$ periods of the potential. 

The Harper model possesses an interesting duality property \cite{aubry80}
which becomes evident by transforming Wannier states $\vert n\rangle$ into
new states
\begin{equation}
\vert k\rangle = L^{-1/2}\sum_n \exp(i2\pi k\beta n)\vert n\rangle
= L^{-1/2}\sum_n \exp\left(i2\pi\frac{kF_{i-1}}{L} n\right)\vert n\rangle\,.
\label{eq:harpertrafo}
\end{equation}
These are eigenstates of the momentum operator to eigenvalues
$kF_{i-1}\,\mathrm{mod}\,F_i$. Neighbouring values of $k$ therefore do not 
imply neighbouring momentum eigenvalues.

With the transformation (\ref{eq:harpertrafo}) one obtains the dual Hamiltonian
\begin{equation}
H=\frac{\lambda}{2}\left[\sum_k\left(\vert k\rangle\langle k+1\vert + 
\vert k+1\rangle\langle k\vert\right) + \frac{4}{\lambda}\sum_k 
\cos\left(2\pi\beta k\right)\vert k\rangle\langle k\vert\right]\,.
\label{eq:hhk}
\end{equation}
By this transformation real and momentum space are interchanged and
the original potential strength $\lambda$ becomes inverted into $4/\lambda$.
Comparison of (\ref{eq:hhx}) and (\ref{eq:hhk}) yields the self-dual point
$\lambda=2$ which separates the parameter regimes of extended and localized
states. In contrast to the nearest neighbour coupling in real space in
(\ref{eq:hhx}), the new Hamiltonian does not couple nearest neighbour momenta.
The physical reason is that scattering by the incommensurate potential may 
change the momentum by a large amount. This coupling to quite different 
momentum values will be of central importance for our reasoning below.

\section{Real and momentum space}
After this discussion, the question arises why these two archetypical models 
behave so differently. In particular, what is the physical reason which allows 
a localization transition at finite value $\lambda=2$ in the Harper model? In 
order to answer this question, we first take a look at the structure of the
wave function $\vert\psi\rangle=\sum_n c_n\vert n\rangle$ expressed in terms
of the Wannier states $\vert n\rangle$, which will provide information about 
the spatial extension of the state. An often used quantity is the inverse 
participation ratio in real space \cite{thoul74,hashi92,mirli00}
\begin{equation}
P_x = \sum_n \vert c_n\vert^4\,.
\label{eq:iprx}
\end{equation}
Provided that $\sum_n\vert c_n\vert^2=1$, the inverse of this quantity
indicates the number of lattice sites over which the wave function is
distributed. A corresponding quantity 
\begin{equation}
P_k = \sum_n \vert d_n\vert^4
\label{eq:iprk}
\end{equation}
can be defined in momentum space, where 
\begin{equation}
d_n=L^{-1/2}\sum_l \exp\left(i2\pi\frac{nl}{L}\right)c_l\,.
\end{equation}

\begin{figure}
\centerline{\includegraphics[width=0.8\textwidth]{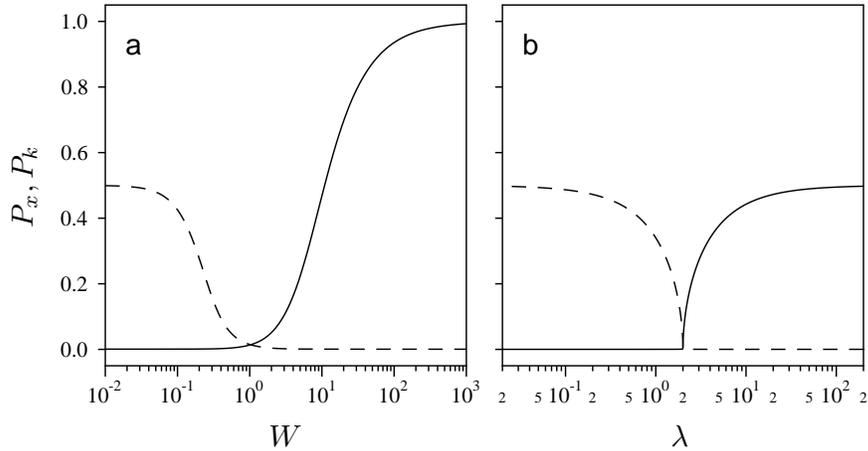}}
\caption{The inverse participation ratio in real space (full line) and
momentum space (dashed line) is shown (a) for the Anderson model with $L=2048$
and (b) for the Harper model with $L=10946$. The curves represent averages over
an ensemble of eigenstates of the respective Hamiltonian as explained in the 
text.}
\label{fig:iprxk}
\end{figure}

The two quantities are depicted as full lines (real space) and dashed lines
(momentum space) in fig.~\ref{fig:iprxk}a for the Anderson model of length
$L=2048$ and fig.~\ref{fig:iprxk}b for the Harper model of length $L=10946$. 
For the Anderson model, the curves represent an average over 50 disorder 
realizations with $L/2$ states around the band center each. For the Harper
model, an average over all symmetric eigenstates has been taken \cite{thoul83}. 

For both models one observes a monotonously increasing inverse participation 
ratio in real space which corresponds to an increasing localization of the
eigenfunctions as the potential strength is increased. Correspondingly, the
inverse participation ratio in momentum space decreases with increasing
potential strength, indicating a delocalization in momentum. The different 
limiting values of $P_x$ for strong potential reflect the fact that in the 
Anderson model the eigenfunctions localize at one site while in the Harper 
model two sites are occupied because we consider here symmetric eigenstates. 

While the overall picture is qualitatively the same for both models, we note
an important difference which becomes apparent when the system size is changed. 
In the Anderson model, the transition from extended to localized states is 
smooth and notably shifts to lower values of $W$ as $L$ is increased. As a 
consequence, in the limit of infinite system size all states are localized if 
finite disorder is present. In contrast, for the Harper model one observes a 
sharp transition at $\lambda=2$ for sufficiently large system sizes.

\section{Phase space approach}
The differences between the two models become much more distinct in phase
space. This approach has the advantage of providing a consistent description 
for arbitrary potential strength \cite{weinm99}. The inverse participation 
ratio in phase space \cite{manfr00,sugit02,wobst02}
\begin{equation}
P=\int\frac{\mathrm{d}x\mathrm{d}k}{2\pi}[\varrho(x,k)]^2
\end{equation}
is based on the positive definite phase space density provided by the Husimi
function \cite{husim40} or Q function \cite{cahil69}
\begin{equation}
\varrho(x_0,k_0) = \vert\langle x_0,k_0\vert\psi\rangle\vert^2.
\end{equation}
Here, the state $\vert\psi\rangle$ is projected onto a minimal uncertainty 
state centered around position $x_0$ and momentum $k_0$. In order to ensure
equal resolution in the two directions of phase space we choose the width
of the Gaussian as $\sigma=\Delta x=\sqrt{L/4\pi}=1/2\Delta k$. 

\begin{figure}
\centerline{\includegraphics[width=0.8\textwidth]{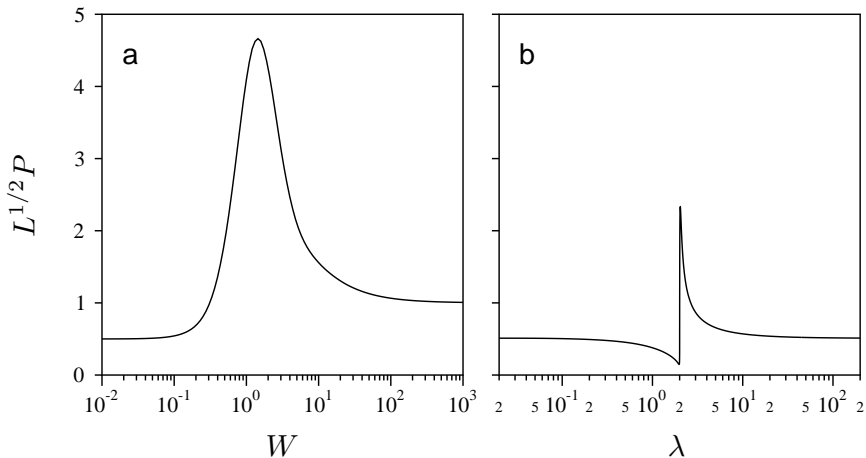}}
\caption{The inverse participation ratio in phase space is shown (a) for the 
Anderson model with $L=2048$ and (b) for the Harper model with $L=10946$.
The averages have been taken with respect to the same states as in
fig.~\protect\ref{fig:iprxk}.}
\label{fig:ipr}
\end{figure}

In fig.~\ref{fig:ipr} we present the inverse participation ratio in phase space
scaled with $L^{1/2}$ which is appropriate in the absence of a potential as
well as for very strong potentials \cite{wobst02}. For the Anderson model 
(fig.~\ref{fig:ipr}a), one obtains an increased inverse participation ratio at 
intermediate potential strengths implying that the eigenstates contract in 
phase space. As we will demonstrate below, the behaviour to the left of the peak 
is dominated by a contraction in real space corresponding to the increase of 
$P_x$ (cf.\ fig.~\ref{fig:iprxk}a) while to the right of the peak the decrease 
is dominated by the decrease of $P_k$. 

The qualitative agreement between the inverse participation ratios in
real and momentum space for the Anderson and Harper model suggests that
the same should hold true for the inverse participation ratio in phase space.
This is even more so since the scenario just described for the Anderson model
is consistent with the duality property of the Harper model where an inversion
of the potential strength is accompanied by a transformation between real and
momentum space. However, the results depicted in fig.~\ref{fig:ipr}b tell a
different story. 

In contrast to the Anderson model, the inverse participation ratio $P$
shown in fig.~\ref{fig:ipr}b initially decreases with increasing potential 
strength up to $\lambda=2$. In this regime, the eigenstates therefore become 
more and more delocalized in phase space. Then, at $\lambda=2$, the phase space 
distribution contracts and starts to become delocalized in momentum as 
$\lambda$ is increased further. Therefore, for almost all values of $\lambda$, 
the phase space behaviour is dominated by the momentum component except for the 
transition which is dominated by the real space behaviour. 

The question now arises, why for weak potential the Anderson and Harper models 
behave so differently while for strong potentials they behave in the same way. 
The mechanism at work for weak potential can be considered to be responsible 
for the localization transition in the Harper model because the jump found in 
fig.~\ref{fig:ipr}b can only occur if the phase space distribution broadens 
as $\lambda$ is increased from zero.

\section{Uncertainty in phase space}
Before addressing this question in detail, we recall the Gaussian smearing 
arising from the projection onto minimal uncertainty states which is inevitable 
if a positive definite phase space density is required. As a consequence, the 
Husimi function only provides limited resolution which, as discussed above, we 
have chosen to be equal for the spatial and momentum components. The resolution 
is of the order of $\sqrt{L}$ sites which in the thermodynamic limit becomes 
small compared to the number $L$ of sites in real as well as momentum space. It 
should be kept in mind, however, that for large $L$ a phenomenon occurring on a 
fixed and finite number of sites cannot be resolved. These considerations would 
still hold, if we chose $\Delta x$ and $1/\Delta k$ to scale according to 
$L^{\alpha}$ with $0<\alpha<1$. In contrast, an entirely different situation
arises, if we keep the resolution fixed in one of the two phase space
directions as is the case for the inverse participation ratios in real and
momentum space, (\ref{eq:iprx}) and (\ref{eq:iprk}). Then, no effects occurring 
in the other direction can be resolved even in the thermodynamic limit.

\section{Limit of strong potential}
We are now in a position to answer the question raised above. It is useful to 
start by considering the limit of strong potential where the phase space 
behaviour of the Anderson and the Harper model agree. For infinitely strong 
potential, the eigenstates of the Anderson model are localized at one site 
while for the Harper model it is sufficient to restrict the discussion to one 
of the two sites occupied by a symmetric eigenstate. The kinetic energy may now 
be considered as a perturbation coupling in lowest order to the two neighbouring 
sites. The dominant contribution will be due to the neighbouring site closer in 
energy so that the problem reduces to the solution of a two-level system 
\cite{wobstxx}. 

The energy eigenstates at very large but finite potential strength will be 
delocalized over two sites. While this will reduce the inverse participation 
ratio $P_x$ in real space, it will not affect the inverse participation ratio 
$P$ in phase space which can only resolve spatial structures of size $\sqrt{L}$
and larger. We recall, that this argument is independent of our particular 
choice of $\sigma$ since the absolute width of the minimal uncertainty state 
has to increase with increasing system size even though its relative width 
decreases to ensure a proper classical limit.

The two original eigenstates which are coupled by means of the nearest-neighbour 
hopping term were localized on one site each and therefore totally delocalized 
in momentum space. Introducing the coupling, one finds a reduced spreading in 
momentum in order to ensure the orthogonality of the two states. This can 
easily be verified by considering the superpositions $(\vert n\rangle\pm\vert
n+1\rangle)/\sqrt{2}$ which in momentum space exhibit large scale density
oscillations of period $L$. By means of the Fourier transformation the effect
of the coupling on short real space distances, too small to be resolved in phase
space, is turned into a large scale phenomenon. Therefore, in the momentum 
component the coupling is easily detected even with the finite resolution of 
the Husimi function. As a consequence, the behaviour of the inverse 
participation ratio in phase space is dominated by momentum and one finds an 
increase of the inverse participation ratio with decreasing potential strength 
as depicted in fig.~\ref{fig:ipr}.

\section{Limit of weak potential}
We now apply similar considerations to the regime of weak potential by starting
from momentum eigenstates, \textit{i.e.}\ states well localized in momentum 
analogous to the localized states in real space considered before. In the
Anderson model the random potential will lead to a coupling among all momentum
eigenstates. However, as before, the coupling between states close in energy,
and therefore close in momentum, will be most effective. As a consequence, the 
role of position and momentum are interchanged and with the same arguments as 
above, we find an increase of the inverse participation ratio in phase space 
with increasing potential strength, albeit now due to the behaviour in real 
space. Within this perturbative treatment, we can readily understand the 
behaviour of $P$ depicted in fig.~\ref{fig:ipr}a.

The situation is quite different for the Harper model where we may consider
the dual model (\ref{eq:hhk}) for small $\lambda$. The perturbation is now
represented by the first term on the right-hand side of (\ref{eq:hhk}) which 
couples to well-defined momentum eigenstates. However, as remarked below 
(\ref{eq:hhk}), due to the scattering by the incommensurate potential these 
eigenstates in general do not correspond to nearest neighbour momenta. For most 
of the energy eigenstates, the momentum eigenstates to which the coupling 
occurs are far away on the scale of the resolution of the Husimi function. The 
resulting broadening of the momentum distribution leads to a reduction of the 
inverse participation ratio. In real space, on the other hand, the coupling 
will lead to short scale oscillations which are not resolved because of the 
finite resolution $\sigma$ in phase space. Therefore, the influence of the 
momentum component dominates and the inverse participation ratio decreases with 
increasing potential strength.

Even in the Harper model there exist few particular states which couple to 
states close in momentum. Then, the inverse participation ratio will initially 
rise. However, the next order coupling leads to a distant momentum value and 
the inverse participation ratio in phase space will eventually decrease before 
reaching $\lambda=2$. 

It follows from this discussion that, in contrast to the Anderson model, the 
Harper model for both weak and strong potential is dominated by the momentum 
properties. It is only around $\lambda=2$ that real space becomes important.
From the comparison of the Anderson and the Harper model we conclude, that the 
form of the coupling between the momentum eigenstates due to a weak potential 
plays a decisive role for the structure of the eigenstates in phase space
and for the appearance of a delocalization-localization transition.

As a further example we briefly comment on the Anderson model in two and three 
dimensions where the inverse participation ratio in phase space behaves very 
much like in the case of the Harper model (cf.\ fig.~\ref{fig:ipr}b) 
\cite{wobst02}. In the marginal case of two dimensions, the tendency towards a 
transition is therefore clearly visible, even though the critical disorder 
strength vanishes in the thermodynamic limit \cite{wobstxx} and no true phase 
transition occurs. In contrast, in three dimensions the Anderson transition is 
recovered \cite{abrah79,lee85}.
The main difference between the Anderson model in one dimension on the one hand
and in two and three dimensions on the other hand lies again in the coupling
of momentum eigenstates by a weak random potential. In the one-dimensional
case, eigenstates close in energy are necessarily close in momentum. In
contrast, in higher dimensions there may exist even energetically degenerate
states far away in momentum, so that they can be resolved in phase space.

\section{Conclusions}
A crucial aspect of our discussion was the finite resolution available in
phase space. This is in strong contrast to the ideal resolution available
with the inverse participation ratio in real or momentum space, albeit only
in one direction of phase space. As a consequence, there is no possibility
to resolve the other direction even in the limit of large system size.
Accepting Heisenberg's uncertainty relation and thus the finite resolution
in phase space allows one to analyze the structure of the eigenstates in
real as well as momentum space and, as was demonstrated above, to obtain 
valuable information about the model of interest.

This is corroborated by a recent observation by Varga \textit{et al.}\ 
\cite{varga02}, that instead of a full-fledged phase space calculation, one can 
alternatively make use of marginal distributions in real and momentum space. 
Even though the inverse participation ratios deduced from them resemble those 
defined in eqs.\ (\ref{eq:iprx}) and (\ref{eq:iprk}) a Gaussian smearing is 
again crucial.

We therefore conclude that for the understanding of the localization properties
of a quantum particle, where both position and momentum are relevant, the
smearing in phase space called for by the uncertainty relation is not only
necessary but also represents an essential ingredient of the physical 
argumentation.

\acknowledgments
The authors thank D.~Weinmann, I.~Varga, C.~Schuster and H.~J.~Korsch for 
interesting discussions. This work was supported by the 
Son\-der\-for\-schungs\-be\-reich 484 of the Deutsche Forschungsgemeinschaft. 
The numerical calculations were carried out partly at the Leibniz-Rechenzentrum 
M{\"u}nchen.

\end{document}